\documentclass[pss,fleqn,embeddedheads]{w-art}
\usepackage{times}
\usepackage{w-thm}
\usepackage[]{graphicx}
\begin{document}
%%    The information for the title page will be placed between
%%    \begin{document} and \maketitle. The order of most entries
%%    is determined by the class file and can not be changed by
%%    rearranging them. The maketitle command follows after the
%%    abstract.
%%
%%    Most of the following commands will be completed by the publisher.
%%
%%    The copyrightyear is defined in the .clo file as the first argument
%%    of the copyrightinfo command. If the copyrightyear differs from that
%%    value it might be adjusted by the following definition:
%%
%% \renewcommand{\copyrightyear}{2003}% uncomment to change the copyrightyear.
%%
\DOIsuffix{theDOIsuffix}
%%
%% issueinfo for header and copyright line
\Volume{XX}
\Issue{1}
\Month{01}
\Year{2003}
%%
%%    First and last pagenumber of the article. If the option
%%    'autolastpage' is set (default) the second argument may be left empty.
\pagespan{1}{}
%%
%%    Dates will be filled in by the publisher. The 'reviseddate' and
%%    'dateposted' (Published online) entry may be left empty.
\Receiveddate{}
\Reviseddate{}
\Accepteddate{}
\Dateposted{}

%% \pretitle{Editor's Choice}

%% We have a short and a long form for the title. The short form
%% (optional argument) goes into the running head.

\title[Interplay of orbitally polarized and magnetically ordered phases]
{Interplay of orbitally polarized and magnetically \\ ordered phases 
in doped transition metal oxides }

%% Please do not enter footnotes or \inst{}-notes into the optional
%% argument of the author command. The optional argument will go into
%% the header.  If there is only one address the marker \inst{x} may be
%% omitted.
                                       
%% Information for the first author.
\author[R. Fr\'esard]{Raymond Fr\'esard\footnote{Corresponding
     author: e-mail: {\sf Raymond.Fresard@ensicaen.fr}, 
     Phone: +33\,231\,452\,609,
     Fax: +33\,231\,951\,600}\inst{1}} 
\address[\inst{1}] {Laboratoire CRISMAT, UMR CNRS-ENSICAEN (ISMRA) 6508,
Caen, France}
%%
%%    Information for the second author
\author[M. Raczkowski]{Marcin Raczkowski\inst{1,2}}
\address[\inst{2}] {Marian Smoluchowski Institute of Physics, Jagellonian
 University, Reymonta 4, 30059 Krak\'ow, Poland}
%%
%%    Information for the third author
\author[A.~M. Ole\'s]{Andrzej M. Ole\'s\inst{2}}
%%
%%    \dedicatory{This is a dedicatory.}

\begin{abstract}
We investigate the magnetic instabilities of the two-dimensional
model of interacting $e_g$ electrons for hole doping away from two
electrons per site in the mean-field approximation. In particular,
we address the occurrence of orbitally polarized states due to the
inequivalent orbitals, and their interplay with ferromagnetic and 
antiferromagnetic spin order. The role played 
by the Hund's exchange coupling $J_H$ and by the crystal field orbital
splitting $E_z$ in stabilizing one of the competing phases is discussed 
in detail.
\end{abstract}

%% maketitle must follow the abstract.
\maketitle                   % Produces the title.

%% If there is not enough space inside the running head
%% for all authors including the title you may provide
%% the leftmark in one of the following three forms:

%% \renewcommand{\leftmark}
%% {First Author: A Short Title}

%% \renewcommand{\leftmark}
%% {First Author and Second Author: A Short Title}

%% \renewcommand{\leftmark}
%% {First Author et al.: A Short Title}

%% \tableofcontents  % Produces the table of contents.

\section{Introduction}

The richness of cooperative phenomena encountered in transition metal
oxides continues attracting considerable attention. In addition to the
spin and charge degrees of freedom, also the coupling to the lattice
and, in particular, the orbital degrees of freedom at and close to the
orbital degeneracy lead to very interesting behavior \cite{Mae04}. Here
we will focus on the phenomena observed in nickelates and manganites, 
which can be attributed to strongly correlated $e_g$ electrons. So far, 
it is known that in the regime of large intraorbital Coulomb interaction 
$U$ strong quantum fluctuations may lead to qualitatively new behavior 
in a Mott insulator with three $e_g$ electrons per site \cite{Fei97}, 
but the competition between various magnetic and orbital instabilities 
was little explored in the weak coupling regime. 

While several features are generic in the models with two \cite{Kle00} 
or more \cite{Flo02} orbitals per ion, and occur already when diagonal 
hopping is assumed, we note that these models are closer to the behavior 
of $t_{2g}$ electrons --- strong interactions between them might explain 
a ferromagnetic (FM) instability in ruthenates \cite{Fre02}. In contrast, 
the orbital flavor for $e_g$ electrons is not conserved, and this is 
likely to lead to partial orbital polarization which is expected to 
modify the magnetic instabilities. Here we will consider such a realistic 
two-dimensional (2D) model of $e_g$ electrons with $(dd\sigma)$ hopping
element $t$ at intermediate and strong coupling, which includes the full 
structure of on-site Coulomb interactions with two parameters: the 
Hubbard element $U$ and Hund's exchange $J_H$ \cite{Amo83}, and the 
crystal field orbital splitting $E_z$.

The magnetic and orbital instabilities within the $e_g$ band were less
investigated until now, but they become very relevant in the context
of doped La$_{2-x}$Sr$_x$NiO$_4$ nickelates, where interesting novel 
phases including the stripe order were discovered \cite{Tra95}.
They were obtained in the theory using realistic models both in 
Hartree-Fock \cite{Miz97} and in exact diagonalization of finite 
clusters including the coupling to the lattice \cite{Hot04}. 
Furthermore, the role of Hund's exchange in the FM instability and in 
the metal-insulator transition was emphasized using a multiband 
Gutzwiller wave-function \cite{Web98}. All these studies reveal 
interesting competition between FM and antiferromagnetic (AF) 
instabilities at density $n=1$, and the competition between the latter 
two and a $C$-AF instability at $n=1.5$. In this paper we investigate 
the pure electronic problem for $e_g$ electrons, using the mean-field 
approximation, seeking for phases which are both orbitally and 
magnetically polarized. By varying the electron density $n$ between
half filling $n=2$ and $n=0$, we cover the hole doping regime $x=n-2$ 
relevant to the nickelates.  

\section{Model}

In this work we consider a two-dimensional (2D) model for $e_g$ 
electrons on a square lattice,
\begin{equation}
{\cal H}=H_{\mathnormal kin} + H_{\mathnormal int} + H_{\mathnormal z},
\label{eq:H}
\end{equation}
with two orbital flavors: $|x\rangle\sim|x^{2}-y^{2}\rangle$ and
$|z\rangle\sim|3z^{2}-r^{2}\rangle$. The kinetic energy is described by
\begin{equation}
 H_{\mathnormal kin}= \sum_{\langle ij\rangle}\sum_{\alpha\beta\sigma}
           t^{\alpha\beta}_{ij}
           c^{\dag}_{i\alpha\sigma}c^{}_{j\beta\sigma}, \qquad
           t^{\alpha\beta}_{ij}=-\frac{t}{4}
\begin{pmatrix}
3           & \pm\sqrt{3} \\
\pm\sqrt{3} &  1
\end{pmatrix},
\label{eq:H_kin}
\end{equation}
where $t$ stands for an effective $(dd\sigma)$ hopping matrix element
due to the hybridization with oxygen orbitals on Ni$-$O$-$Ni bonds, and 
the off-diagonal hopping $t^{xz}_{ij}$ along $a$ and $b$ axis depends on 
the phase of the $|x\rangle$ orbital along the considered cubic direction. 
The electron-electron interactions are described by the on-site terms, 
which we write in the following form \cite{Amo83}, 
\begin{eqnarray}
 H_{\mathnormal{int}} &=&
       U\sum_{i}\bigl( n^{}_{ix\uparrow}n^{}_{ix\downarrow}
                    +  n^{}_{iz\uparrow}n^{}_{iz\downarrow}\bigr )
    + \bigl(U-\tfrac{5}{2}J_H\bigr)\sum_{i}n^{}_{ix}n^{}_{iz} \nonumber \\
   &-& 2J_H\sum_{i}\textbf{S}_{ix}\cdot\textbf{S}_{iz} 
    + J_H\sum_{i}\bigl( c^{\dag}_{ix\uparrow}c^{\dag}_{ix\downarrow}
                        c^{}_{iz\downarrow}c^{}_{iz\uparrow} 
                      + c^{\dag}_{iz\uparrow}c^{\dag}_{iz\downarrow}
                        c^{}_{ix\downarrow}c^{}_{ix\uparrow} \bigr),
\label{eq:H_int}
\end{eqnarray}
with $n^{}_{i\alpha}=\sum_{\sigma}n_{i\alpha\sigma}$ (for $\alpha=x,z$).
$U$ and $J_H$ stand for the intraorbital Coulomb and Hund's 
exchange elements. The interactions $H_{\mathnormal{int}}$ are
rotationally invariant both in the spin and in the orbital space. 
The last term $H_{\mathnormal z}$ describes the uniform crystal-field 
splitting between $x^2-y^2$ and $3z^2-r^2$ orbitals,
\begin{equation}
 H_{\mathnormal z} = \tfrac{1}{2}E_z\sum_{i} (n_{ix}-n_{iz}).
\label{eq:H_z}
\end{equation}

It is convenient to rewrite the Hamiltonian
(\ref{eq:H_int}) by introducing the following operators:
\begin{eqnarray}
\label{eq:n}
n_i &=& \sum_{\alpha\sigma}n_{i\alpha\sigma},                  \\
\label{eq:op}
m_i&=&\sum_{\alpha\sigma}\lambda_{\sigma}n_{i\alpha\sigma}, \hskip .7cm
o_i = \sum_{\alpha\sigma}\lambda_{\alpha}n_{i\alpha\sigma}, \hskip .7cm
p_i = \sum_{\alpha\sigma}\lambda_{\alpha}
                          \lambda_{\sigma} n_{i\alpha\sigma},  \\
\label{eq:fluct}
f_{i\sigma} &=& \sum_{\alpha\beta} c^{\dagger}_{i\alpha\sigma}
\sigma^x_{\alpha\beta} c^{}_{i\beta\sigma}, 
\end{eqnarray}
with $\lambda_{\sigma}=\pm 1$ for $\sigma=\uparrow(\downarrow)$ spin 
and $\lambda_{\alpha}=\pm 1$ for $\alpha=x(z)$ orbital, and $\sigma^x$ 
is a Pauli matrix. These operators 
correspond to the total density, the total magnetization, the orbital 
polarization, the magnetic orbital polarization, and the on-site
orbital flip respectively. The Coulomb interaction term
$H_{\mathnormal{int}}$ (\ref{eq:H_int}) can be then written as:
\begin{eqnarray}
H_{\mathnormal{int}} &=& \tfrac{1}{8}\sum_{i}\big[
(3U-5J_H) n_{i}^2-(U+J_H) m_{i}^2-(U-5J_H) o_{i}^2-(U-J_H)p_{i}^2\big]      
\nonumber \\
&+& J_H \sum_{i} f_{i\uparrow} f_{i\downarrow}\; .
\label{eq:H_int2}
\end{eqnarray}
 
The order parameters introduced in Eqs. (\ref{eq:op}): $m_i$, $o_i$, and
$p_i$, used next to minimize the ground state energy, provide a complete 
description of the ground state at finite doping. We emphasize that
they also reveal the dominating role of the kinetic energy of doped holes
over the superexchange energy. Namely, large electron filling of 
$|z\rangle$ orbitals, contributing to a narrow band, optimizes the
kinetic energy of holes in magnetically polarized states, doped into the 
$|x\rangle$ orbitals, contributing to a wide band. On the contrary, the
superexchange $\propto J=4(t^{\alpha\alpha})^2/U$ at large $U$ suggests 
that the system would better optimize the magnetic energy when the 
orbitals with larger hopping elements $t^{xx}$ are closer to half 
filling. We show below that the complex interplay between all the degrees 
of freedom of the model (\ref{eq:H}) results in rather peculiar doping 
dependence of the order parameters (\ref{eq:op}), 
and thus leads to highly nontrivial and rich phase diagrams. 
  
We investigated the stability of possible phases with either uniform or
staggered magnetic order in the mean-field (Hartree) approximation by 
expressing the local operators (\ref{eq:op}) by their mean-field 
averages,
$\gamma_i^2\simeq 2\gamma_i\langle\gamma_i\rangle-\langle\gamma_i\rangle^2$.
In order to establish unbiased results, the calculations were carried out 
on a large 128$\times$128 cluster, using periodic boundary conditions 
at low temperature $T=0.01t$ (here $k_B=1$). Consistently with the present 
mean-field analysis we assumed $\langle f_{i\sigma}\rangle=0$.

\section{Numerical results}

\begin{figure}[t!]
\includegraphics[width=.88\textwidth]{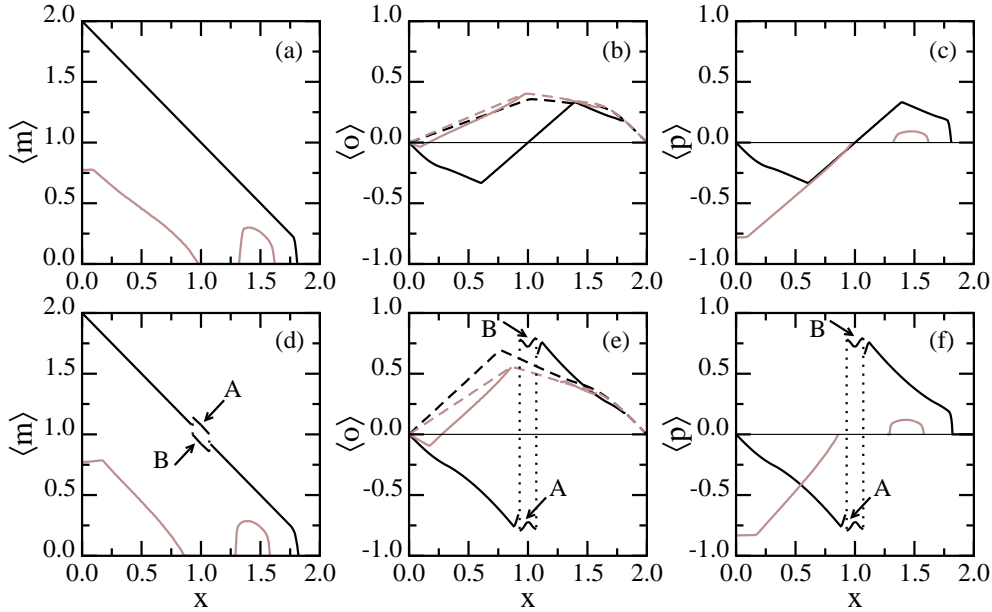}
\caption
{
Order parameters: magnetization $\langle m\rangle$, orbital polarization
$\langle o\rangle$, and magnetic polarization $\langle p\rangle$ in 
the FM phase as a function of doping $x$ for:
(a)-(c) $J_H=0.25U$ and (d)-(f) $J_H=0.15U$, and for two values of the
Stoner parameter: $U+J_H=8t$ (black solid lines) and 
                  $U+J_H=4t$ (gray solid lines).
$A$ and $B$ refer to two sublattices in the orbital ordered state for
$x\sim 1$. The orbital polarization in the reference PM states is shown 
by dashed lines.
} 
\label{fig:FM}
\end{figure}

\subsection{\it Magnetic order and orbital polarization:}

First of all, in the paramagnetic (PM) state at $E_z=0$, with 
$\langle m\rangle=\langle p\rangle=0$, a higher electron density is found 
in $|x\rangle$ orbitals ($\langle o\rangle>0$), as then the kinetic 
energy is lowered [Figs.~\ref{fig:FM}(b) and \ref{fig:FM}(e)], except at 
$x=0$. This state is our reference state for possible magnetic 
instabilities.    

We now proceed with the discussion of the magnetic order and orbital
polarization for two characteristic values of the Stoner parameter 
$I \equiv U+J_H$: intermediate coupling $I=4t$, and strong coupling 
$I=8t$, being smaller and larger than the bandwidth $W=6t$, respectively.
Let us begin with the FM phase. In Fig.~\ref{fig:FM}(a) we show the 
magnetization $\langle m\rangle$ as a function of doping $x$ for the 
ratio $J_H/U=0.25$ which is representative of the strong Hund's 
exchange coupling regime. In this case the interaction in the 
$o$-channel is \textit{repulsive}. As shown in Fig.~\ref{fig:FM}, 
several FM phases occur. 

Consider first the intermediate interaction strength $I=4t$
[Figs.~\ref{fig:FM}(a-c)], where one finds two disconnected FM states: 
one for $0\leq x<1$, and the second one for $x\simeq 1.5$. The latter 
corresponds to a van Hove singularity in the density of states. Since 
it is predominantly related to the $|x\rangle$ orbital, both the orbital 
polarization $\langle o\rangle$ [Fig.~\ref{fig:FM}(b)] and the magnetic 
polarization $\langle p\rangle$ [Fig.~\ref{fig:FM}(c)] are positive in 
this doping regime. In contrast, for $0\leq x<1$ the total energy is 
minimized when a higher electron density is found in the $|z\rangle$ 
orbital, where also larger magnetic moments are formed. Such anisotropic 
filling of $e_g$ orbitals follows from a large difference between the 
$t^{xx}$ and $t^{zz}$ hopping elements \cite{Zaa93}. Note that the 
orbital polarization is here opposite to that in the reference PM state.  

The above peculiar behavior disappears gradually when the interaction 
strength is enhanced to $I=8t$ and both $\langle o\rangle$ and 
$\langle p\rangle$ tend to saturate to the optimal value, being positive 
(negative) for $x<1$ ($x>1$), leaving the $|z\rangle$ orbital almost
fully polarized. Here the magnetic instability, with the largest
effective interaction term $U+J_H$, dominates and the magnetization 
barely deviates from its saturation value $2-x$, except for the low 
electron density $x>1.8$. In this case doping the half-filled FM state 
first leads to holes introduced into the $|x\rangle$ orbitals, leaving 
the center of the narrower band, with predominantly $|z\rangle$ orbital 
character, below the former broader one. Therefore, in this situation 
the formation of localized magnetic moments optimizes the energy.
They are naturally associated with the $|z\rangle$ orbitals since 
they contribute to the narrower band. 

When Hund's exchange coupling $J_H$ is reduced, the interaction in 
the $o$-channel becomes \textit{attractive}. As a result, for large 
$I=8t$, both $\langle o\rangle$ and $\langle p\rangle$ nearly 
saturate to the ideal behaviors $-x$ for $x<1$ and $2-x$ for $x>1$.
Therefore, a transition between these two solutions would be first 
order, and one observes a jump at $x\sim 1$. However, as shown in 
Figs.~\ref{fig:FM}(d-f), the two-sublattice orbital order sets in in 
this crossover regime in the form of a FM$_{xz}$ state. This state has  
opposite orbital polarization $\langle o\rangle_A\sim -0.8$ and 
$\langle o\rangle_B\sim 0.8$ on both sublattices [Fig.~\ref{fig:FM}(e)]. 
While the total density $\langle n\rangle_A$ is somewhat higher than 
$\langle n\rangle_B$ due to inequivalent $e_g$ orbitals, also 
$\langle m\rangle_A>\langle m\rangle_B$. For large doping $x>1$ the 
electrons occupy mainly the $|x\rangle$ orbitals, and the small 
occupancy of $|z\rangle$ orbital results solely from the interorbital 
hopping term $\propto t_{xz}$. Indeed, at $x\sim 1.3$ one finds 
appreciable orbital polarization, with $|x\rangle$ orbitals 
occupied and almost empty $|z\rangle$ orbitals, the situation 
encountered in La$_{1-x}$Sr$_{1+x}$MnO$_4$ manganites \cite{Mac99}. 
In all FM phases found at $J_H=0.15U$ the total magnetization is close 
to saturation. When $U$ is reduced, one gradually recovers the behavior 
obtained for large $J_H/U$.
 
\begin{figure}[t!]
\includegraphics[width=.88\textwidth]{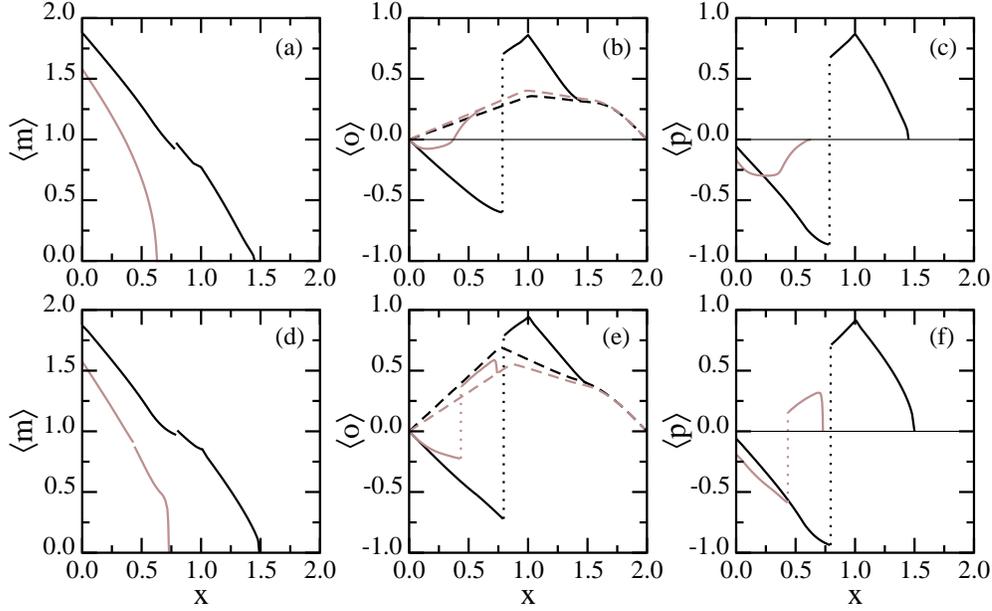}
\caption
{
Order parameters as in Fig.~\ref{fig:FM} but for the AF phase.
} 
\label{fig:AF}
\end{figure}

We now turn to the AF phase, expected as a ground state near half
filling ($x=0$). The order parameters are shown in Fig.~\ref{fig:AF} 
for the same parameter values as for the FM case. For $I=8t$ and 
$J_H/U=0.25$ the mean-field equations possess two competing solutions 
[Figs.~\ref{fig:AF}(a-c)]. The first one, which can be continued to 
weak coupling, is characterized by negative values of orbital 
$\langle o\rangle$ and magnetic $\langle p\rangle$ polarizations. Namely, 
the higher electron density is found within the $|z\rangle$ orbitals, and 
these orbitals carry the magnetic moment. More precisely, introducing 
holes in the half-filled insulating AF state mostly affects the band 
with $|x\rangle$ orbital character, leaving the localized magnetic 
moments within the $|z\rangle$ orbital almost saturated. This solution 
extends to large doping $x\simeq 1$. In contrast, the second solution 
rather stems from the behavior expected for low density: the electrons 
first occupy the broader band with $|x\rangle$ orbital character until 
quarter filling ($x=1$) is reached, and next they gradually occupy the 
other band, with $|z\rangle$ orbital character. However, since the 
interaction in the $o$-channel is \textit{repulsive} and since both 
bands are coupled, the orbital polarization $\langle o\rangle$ and the
magnetic polarization $\langle p\rangle$ are reduced from their maximal 
values which would be reached for decoupled orbitals. 
These two trends from $x=0$ and $x=2$ contradict each other, and 
therefore an abrupt (first order) transition between both solutions is 
observed at $x\simeq 0.7$ [Figs.~\ref{fig:AF}(b-c)].

Reducing $J_H$ barely affects the above findings for strong coupling 
$I=8t$ [Fig.~\ref{fig:AF}(d-f)]. While the magnetization 
$\langle m\rangle$ is almost unchanged, the first order transition 
between two differently polarized states is more pronounced, as the 
values of the orbital polarization and the magnetic polarization are 
enhanced. When reducing $U$ the location of the first order phase 
transition shifts towards smaller doping, while all order parameters
are suppressed. At the same time the critical doping locating the 
second order phase transition is reduced by a weaker Coulomb 
interaction $U$ but is enhanced by a weaker Hund's exchange coupling 
$J_H$. When seeking for other phases one may expect that two-sublattice 
FM solutions can smoothly interpolate between the FM and AF states. 
Such solutions never turned out to be the ground state in this study 
(up to $I=8t$). 

\begin{figure}[t!]
\includegraphics[width=.88\textwidth]{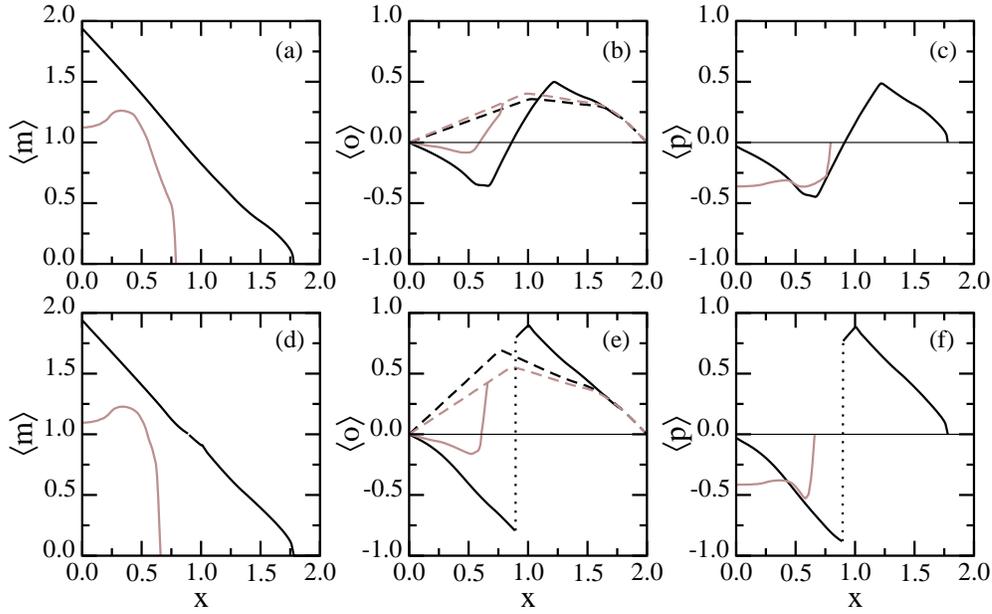}
\caption
{
Order parameters as in Fig.~\ref{fig:FM} but for the $C$-AF phase.
} 
\label{fig:CAF}
\end{figure}

A competition between the FM and AF order in the present model of 
$e_g$ band may lead to a superposition of the two phases in a form
of $C$-AF phase, where the magnetic moments are FM along one direction 
and staggered in the other (orthogonal) one. According to recent 
numerical simulations \cite{Hot04}, a coexistence of FM and AF bonds 
is indeed expected for $x\simeq 0.5$. Unlike in the AF phase, the order 
parameters are continuous functions of doping for $J_H/U=0.25$, as can 
be seen in Figs.~\ref{fig:CAF}(a-c). This behavior is similar to that 
of the FM case (Fig.~\ref{fig:FM}). Its origin can be attributed to the 
orbital polarization which is substantially stronger in the AF case to 
the extent that is exceeds a certain threshold above which no smooth 
solution can interpolate between the small and large doping regimes. 
For $x\sim 0.5$ the magnetic moment $\langle m\rangle$ is carried by 
the $|z\rangle$ orbital for large $U$, while $\langle m\rangle$ 
decreases and $\langle o\rangle$ changes sign for small $U$.

When reducing $J_H/U$, the orbital polarization is enhanced and a first 
order transition appears for $I=8t$ [see Figs.~\ref{fig:CAF}(d-f)]. 
In this case both the orbital polarization and the magnetic moment are 
predominantly carried by the stronger correlated $|z\rangle$ orbital 
(with a weaker hopping and thus larger ratio $U/t^{zz}$ than $U/t^{xx}$) 
in the physically relevant doping range centered around $x=0.5$. 

\subsection{\it Magnetic phase diagrams:}

\begin{figure}[t!]
\includegraphics[width=.91\textwidth]{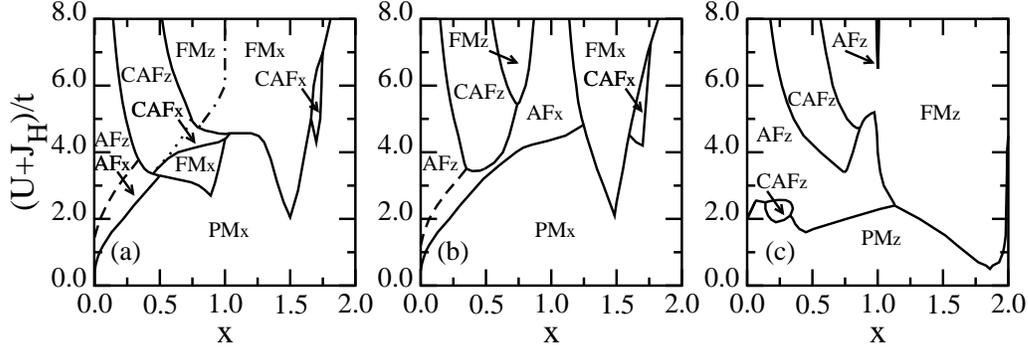}
\caption
{
Phase diagrams of the $e_g$ orbital model (\ref{eq:H}) as functions of 
the Stoner parameter $U+J_H$ and hole doping $x=2-n$, with: 
(a) $J_H=0.25U$, and 
(b) $J_H=0.15U$. Panel (c) shows the stable phases for finite crystal 
field splitting $E_z=2t$ and $J_H=0.25U$. Transitions from the PM phase 
to magnetic phases are second order. The remaining solid lines denote 
first order transitions while the dashed, dotted, and dashed-dotted
lines indicate second order transitions.   
} 
\label{fig:dm}
\end{figure}

Our main findings are summarized in the phase diagrams in 
Fig.~\ref{fig:dm}. For $J_H/U=0.25$ [Fig.~\ref{fig:dm}(a)], the doped 
PM phase is characterized by a positive orbital polarization, therefore 
denoted PM$x$. It is unstable towards AF$x$ 
phase for small doping up to $x\simeq 0.5$, towards $C$-AF$x$ phase for 
$1\leq x\leq 1.02$ and for $1.65\leq x\leq 1.75$, and towards FM$x$ 
phase otherwise. In particular, for $x=1$ the FM$x$, $C$-AF$x$, 
orbitally unpolarized FM and (for $I>16t$) the alternating FM${xz}$ 
phases appear successively with increasing interaction strength $I$.
The $C$-AF phases are found for $x\simeq 0.5$ at $I>4t$, and also in 
a small range around $x\simeq 7/4$.
 
When reducing $J_H/U$ [Fig.~\ref{fig:dm} (b)], the main difference 
appears for $x\sim 1$. Here the PM$x$ phase is unstable towards an AF$x$ 
phase which itself is robust and remains stable up to strong coupling. 
This seemingly peculiar behavior can be better understood by 
diagonalizing exactly a two-site cluster. One can find the ground state 
to be AF$x$ for small $J_H$ and arbitrary $U$ and FM${xz}$ for large 
$J_H/U$, in qualitative agreement with our mean-field calculation. 

Let us finally mention that the model we use is known to have a van Hove
singularity in the vicinity of $x=1.5$, which is expected to induce 
a FM instability for arbitrary weak coupling at zero temperature. This 
particular instability, however, turns out to be unusually strongly 
temperature dependent. Therefore, the corresponding critical value of 
the Stoner parameter $I$ is finite at temperature $T>0$, and reaches 
a value close to $W/3$ (Fig. \ref{fig:dm}) at $T=W/600$ used in this work.  

\subsection{\it Consequences of the crystal field splitting:}

A complete investigation of the phase diagrams at finite crystal field 
splitting $E_z$ would be quite involved, and is left for future work. 
At $E_z=0$ the majority of stable magnetic solutions is characterized by 
a positive orbital polarization, a tendency expected for a 2D model of
$e_g$ electrons \cite{Mac99}, which would certainly be enhanced by a 
negative crystal field $E_z$. We therefore limit our present discussion 
to the influence of a positive $E_z$, in order to investigate a 
competition between the kinetic energy, which is lower when the broad 
band with predominantly $|x\rangle$ orbital character is closer to half
filling, and the potential energy at finite $E_z$. Furthermore, we 
restrict ourselves to the strong Hund's exchange coupling regime 
$J_H=0.25U$.

As shown in Fig.~\ref{fig:dm}(c), the orbital polarization of the PM 
phase is changed to negative (PM$z$ phase) already for moderate $E_z=2t$. 
As a result, the magnetic moment and the orbital polarization are carried 
by the same orbital in all phases, and the magnetic instabilities are 
achieved for lower values of $I=U+J_H$. Another consequence of  
finite $E_z>0$ is the observed shift of the van Hove singularity to 
larger doping, strongly enhancing the tendency towards ferromagnetism in 
the low density regime. In addition, the competition between FM and AF
phases at quarter filling ($x=1$) remains quite spectacular: even though 
both weak coupling and strong coupling expansions predict 
antiferromagnetism, as does our calculation, FM phase nevertheless 
takes over in an intermediate coupling regime $5.15t < I < 6.5t$.

Such features as the AF$z$ phase obtained for low doping, the $C$-AF$z$ 
phase for $x\sim 0.5$, and FM one for both $\sim 0.5<x<0.98$ 
and $x>1.02$, are robust and are found also at finite crystal field, 
in particular in the strong coupling regime. On the contrary, the 
instability of the PM phase to the FM$x$ phase disappears, and the one 
to the $C$-AF phase moves from $x\sim 1.5$ to small doping.

\section{Conclusions}

In summary, we have determined the phase diagram of $e_g$ electrons on 
the square lattice within the mean-field approximation. The occurrence 
of antiferromagnetism in the vicinity of half filling, followed by 
$C$-AF phase at $x\simeq 0.5$, as well as FM phases for $x\simeq 0.75$ 
and $x\simeq 1.5$, are robust features of this model. Note that the 
regions of stability of the AF and $C$-AF phases with respect to the FM 
one would still be somewhat extended due to quantum corrections 
\cite{Rac02}. In particular, the occurrence of $C$-AF phase indicates 
that even more complex types of magnetic phases, such as stripe phases 
with larger magnetic unit cells \cite{Tra95,Hot04}, might be expected 
in doped nickelates.

In contrast, the ground state is strongly parameter dependent in the 
vicinity of quarter filling, $x=1$, resembling to some extent the 
ground state of the model with two equivalent orbitals \cite{Kle00,Fre02}. 
While the orbital polarization systematically appears in all phases, 
the orbital carrying the magnetic moment does not necessarily coincide 
with the one carrying higher electron density, leading to a particularly 
interesting interplay between magnetic and orbital degrees of freedom.

\begin{acknowledgement}
M. Raczkowski was supported by a Marie Curie fellowship of the
European Community program under number HPMT2000-141.
A. M. Ole\'s would like to acknowledge support 
by the Polish State Committee of Scientific Research (KBN) under 
Project No. 1~P03B~068~26.
\end{acknowledgement}

\end{document}